\documentclass[prb,twocolumn,aps,superscriptaddress]{revtex4-1}

\usepackage{graphicx}
\usepackage{dcolumn}
\usepackage{bm}
\usepackage{amssymb}
\usepackage{amsmath}
\usepackage{subfigure}
\usepackage{physics}
\usepackage{float}
\usepackage{hyperref}

\begin{document}

\title{Thermoelectric figure of merit enhancement in dissipative superlattice structures}

\author{Pankaj Priyadarshi}
 \email{priyadarshi56@gmail.com}
\author{Bhaskaran Muralidharan}
 \email{bm@ee.iitb.ac.in}

\affiliation{Department of Electrical Engineering \\
	Indian Institute of Technology Bombay, Powai, Mumbai- 400076, India \\}


\begin{abstract}
Utilizing the non-coherent quantum transport formalism, we investigate thermoelectric performance across dissipative superlattice configurations in the linear regime of operation. Using the {\it{dissipative}} non-equilibrium Green’s function formalism coupled self-consistently with the Poisson's equation, we report an enhanced figure of merit $zT$ in the multi-barrier device designs. The proposed enhancement, we show, is a result of a drastic reduction in the electronic thermal conductance triggered via non-coherent transport. We show that a maximum $zT$ value of 18 can be achieved via the inclusion of non-coherent elastic scattering processes. There is also a reasonable enhancement in the Seebeck coefficient, with a maximum of $1000~\mu V/K$, which we attribute to an enhancement in electronic filtering arising from the non-coherent transport. Distinctly the thermal conduction is drastically reduced as the length of the superlattice scales up, although the power factor shows an overall degradation. While the presence of interfaces is known to kill phonon thermal conduction, our analysis shows that non-coherent processes in superlattice structures can effectively kill electronic thermal conduction also. We believe that the analysis presented here could set the stage to understand better the interplay between non-coherent scattering and coherent quantum processes in the electronic engineering of heterostructure thermoelectric devices.
\end{abstract}
\maketitle
\section{Introduction}
Thermoelectric (TE) materials, structures or devices are typically assessed via the dimensionless figure of merit $zT=S^2 \sigma / (\kappa_{el} + \kappa_{ph})$ that is directly related to its energy conversion efficiency. Here $S$ is the Seebeck coefficient (thermopower), $\sigma$ the electrical conductivity, $\kappa_{el(ph)}$ the electronic (phonon) contribution to the thermal conductivity and $T$ the operating temperature \cite{Ioffe1956, Rowe2005, Goldsmid2010}. In the past two and a half decades, a great deal of research has been carried out in pursuit of an increase in ‘$zT$’, from the conventional bulk to nanostructured thermoelectric materials \cite{Majumdar2004, Snyder2008, Minnich2009}.  \\
\indent Using nanostructuring concepts, a higher value of $zT$ was shown to be achieved by either increasing the numerator or decreasing the denominator part of the figure of merit definition. The numerator is called the power factor (PF $= S^2\sigma$), which mainly constitutes electronic transport in the ballistic or diffusive limit \cite{Fisher2013}. The denominator part is a combination of thermal conduction due to electrons as well as phonons, so there is a trade-off between the PF and the thermal conductivity \cite{MinnichPhDThesis,  Muralidharan2012, Bitan2016, Bitan2018,De_2019}.  An important direction enroute, apart from engineering the reduction of phonon thermal conductivity via enhanced scattering at nano-interfaces \cite{Minnich2009,Aniket2015,Aniket2017}, has been in the direction of low-dimensional nanostructuring to enhance the electronic figure of merit via a distortion in the electronic density of states \cite{Hicks1993-1, Hicks1993-2, Hicks1996, Heremans2013, Broido1995,Balandin2003, Kim2009, Kim2011,Mao2016}.\\ 
\indent As an ultimate example of electronic engineering, it was mathematically proposed that a Dirac-delta transport function can lead to an infinite electronic figure of merit and a conversion efficiency tending to the Carnot limit \cite{Sofo-Mahan1996}. But, the electrical power drawn at this efficiency would be zero in accordance with reversible thermodynamics, whereas, the ideal thermoelectric effect is an irreversible process accompanied by a finite power generation \cite{Nakpathomkun2010, Muralidharan2012, Sothmann2013} . In this context, it was clearly postulated \cite{Whitney2014, Whitney2015} that under the situation of drawing the maximum power at a given efficiency, the ideal electronic transmission would assume a 'box car' or a band pass filtering shape.  In this context, a superlattice structure fits very well where mini-bands can be engineered to assume the box car transmission function. \\
\indent While the idea to use superlattice (SL) structures in thermoelectrics was proposed to enable the suppression of the cross-plane thermal conduction without affecting the condition that enhances the electronic power factor \cite{YangGang2018, Xiao2018}, the idea of using mini-bands in SL structures was also recently proposed in various theoretical efforts\cite{Karbaschi2016,Swarnadip2018,Pankaj2018,myPRA}. It is noted that the phonon mean free path is greater than the size of nanostructures, while the electron mean free path is smaller \cite{Fisher2013, WangWang2014}, giving plenty of room for electronic engineering \cite{LNEDatta}. While the previous works \cite{Karbaschi2016,Swarnadip2018,Pankaj2018,myPRA} focused on the analysis in the ballistic regime, we advance the framework to examine various superlattice structures  depicted in Fig.~\ref{Device}  in the dissipative regime \cite{Aniket2017,Aniket_Peltier}, in the presence of electron phonon interactions treated in the form of B\"uttiker probes \cite{QTDatta,LNEDatta}. Most importantly, the systematic introduction of scattering processes puts us into an interesting perspective where the effects of coherent processes like tunneling and mini-band formation compete with the relaxation processes induced via scattering \cite{Aniket2017,Aniket_Peltier}.\\
\indent In the realm of proposed SL structures, electronic analogs of the optical concept of anti-reflective layers \cite{Pacher2001, Morozov2002, Pankaj2018, Swarnadip2018}  have also been proposed recently and analyzed in detail as potential candidates for enhanced TE performance.  In this paper, we consider normal and anti-reflective superlattice structures \cite{Pacher2001, Morozov2002, Pankaj2018, Swarnadip2018} that are optimally configured in order to analyze the TE properties. We use the miniband transmission feature of the superlattice to set the device Fermi energy level for positive TE transport \cite{Pankaj2018, Pankaj2019, Pankaj2020_AIP}. Using the non-equilibrium Green’s function(NEGF) formalism coupled self-consistently with the Poisson's equation, we report an enhanced figure of merit $zT$ in the multi-barrier device designs. The proposed enhancement is a result of a drastic reduction in the electronic thermal conductance triggered via non-coherent transport. We show that a maximum $zT$ value of 18 can been achieved via the inclusion of non-coherent elastic scattering processes in the electron transport. There is also a reasonable enhancement in the Seebeck coefficient, with a maximum of $1000~\mu V/K$, which we attribute to an enhancement in electronic filtering arising from the non-coherent transport. Distinctly the thermal conduction is drastically reduced as the length of the superlattice scales up, although the power factor shows an overall degradation. While the presence of interfaces is known to kill phonon thermal conduction, our analysis shows that non-coherent processes in superlattice structures can effectively kill electronic thermal conduction also. We believe that the analysis presented here could set the stage to understand better the interplay between non-coherent scattering and coherent quantum processes in the electronic engineering of heterostructure thermoelectric devices.
\section{Superlattice Structure and Computational Formalism}
\begin{figure}
	\centering
	\includegraphics[height=0.25\textwidth,width=0.4\textwidth]{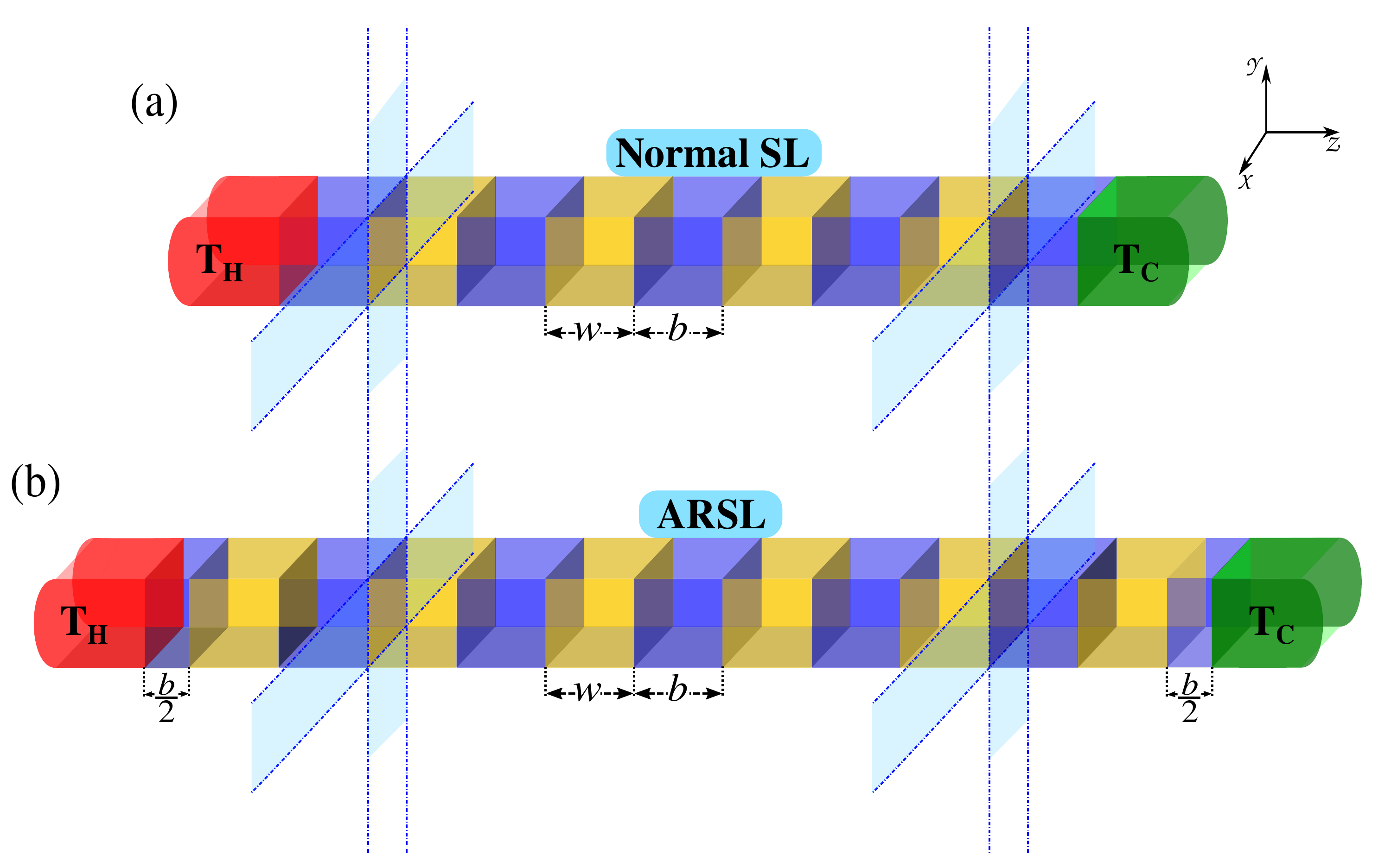}
	\caption{Device schematics: A  finite superlattice structure in the central channel region is sandwiched between the two contacts that maintained at two different temperatures $T_H$ (hot) and $T_C$ (cold), respectively. Here, $w$ is the width of the well region, and $b$ is the barrier thickness. (a) Normal SL is the regular heterojunction structure having a constant well width and barrier thickness. (b) The AR-SL is the anti-reflection enabled SL, in which two additional barriers of half the thickness of a regular barrier is attached after a well width at both the ends.}
	\label{Device}
\end{figure}

Figure~\ref{Device} shows the schematic of a heterostructure thermoelectric device set up, where the two contacts are maintained at temperatures $T_H$ (hot) and $T_C$ (cold), respectively. For building the heterostructure, we choose the GaAs/AlGaAs material system because the lattice constant and the conduction band effective masses are almost the same at small proportion of Al in GaAs. The central channel region of the superlattice configuration in Fig.~\ref{Device}(a), depicts an alternate well-barrier structure in the transport $(\hat{z})$ direction, labeled as Normal SL. The overall device size depends on the number of barriers used, where one unit consist of a barrier of thickness $b$ and a well region of width $w$, successively. Similarly, Fig.~\ref{Device}(b), labeled ARSL (anti-reflective SL), depicts the normal superlattice (NSL) configuration in addition with the two extra half barriers ($b/2$) or the anti-reflective barriers are attached at both the ends after a well region \cite{Pacher2001,Morozov2002,Pankaj2018,Swarnadip2018}. 
We perform a one dimensional carrier transport simulations along the $\hat{z}$ direction in the described SL configurations. To calculate the thermoelectric parameters, we operate our devices in the linear response regime. In this scenario, the thermoelectric dimensionless figure of merit for electronic transport only is calculated as
\begin{equation}
zT_{el} = \frac{S^2 G}{G_K} T_{avg},
\label{eqzT}
\end{equation}
where $S$ being the Seebeck coefficient, $G$, and $G_K$ are the electrical and thermal conductances across the device. Here, $T_{avg}$ is the average temperature of the hot ($T_H$) and cold ($T_C$) contacts. To calculate the quantities in \eqref{eqzT}, we invoke the coupled charge and heat current equations in the linear regime of operation \cite{LNEDatta, Kim2011}, given as
\begin{equation}
I=G\Delta V + G_S\Delta T,
\label{eqI}
\end{equation}
and
\begin{equation}
I_Q = G_P\Delta V + G_Q\Delta T.
\label{eqIQ}
\end{equation}
\begin{figure}
	\centering
	\includegraphics[height=0.35\textwidth,width=0.3\textwidth]{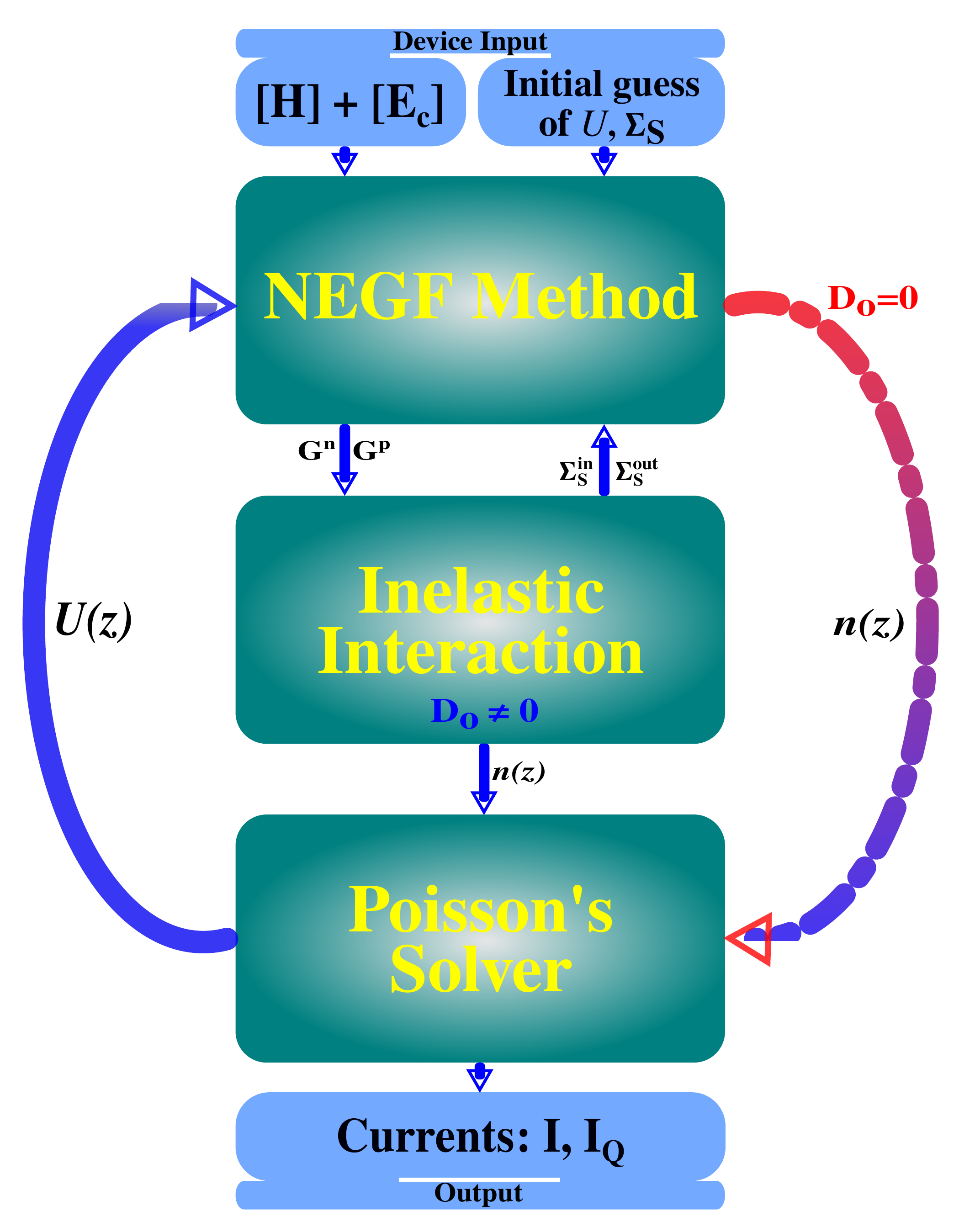}
	\caption{Simulation flow: The blocks depict the simulation formalism starting with the input device Hamiltonian and potential profile. The calculation takes the initial guess of voltage $U$ and the scattering matrix $\Sigma_S$ as input. In the representation, there are two options:- $D_0=0$ (coherent transport) and $D_0 \neq 0$ (non-coherent transport). To solve the transport equation coherently, the bigger loop indicating $D_0=0$ is active, in which the electron density ($n$) is directly provided to the Poisson's solver after the NEGF block. On the other, if $D_0 \neq0$, first, the NEGF and the interaction blocks are solved self-consistently and thus provide the electron density ($n$) to the Poisson's solver. In both cases, the changing  $U$ after the Poisson's block is fed again to the NEGF calculation and thus simulation runs till all the guessed quantities are self-consistently converged.}
	\label{Simulation_Flow}
\end{figure}
\indent In general, the quantities $G$, $G_S$, $G_P$, $G_Q$ are related to the Onsager coefficients \cite{LNEDatta}, under an applied bias  $\Delta V$ and a temperature difference $\Delta T$. \\
\indent To calculate these transport coefficients, we employ the dissipative non-equilibrium Green's function (NEGF) formalism self-consistent with the Poisson charging effect \cite{QTDatta}, elaborated in Fig.~\ref{Simulation_Flow}. The simulation starts with the formation of device Hamiltonian $[H]$ obtained herewith the nearest neighbor tight-binding model. The variation in SL configurations is captured in the form of device potential profile $E_c(z)$. The general dissipative model of quantum transport starts with the energy-resolved retarded matrix representation of the retarded Green's function $[G(E)]$, given by
\begin{equation}
[G(E)]=[E\mathbb{I}-H-E_c-U-\Sigma_H-\Sigma_C-\Sigma_S]^{-1},
\label{eqGS}
\end{equation}
where $E$ is the energy of the electronic wave function, $\mathbb{I}$ is the identity matrix, $\Sigma_H$, and $\Sigma_C$ are the self-energy matrices of hot and cold contacts, respectively. For the electron-phonon interaction, the scattering matrix $[\Sigma_S]$ is calculated as
\begin{equation}
[\Sigma_S] = -\iota \frac{[\Gamma_S]}{2} = \iota \frac{[\Sigma_S^{in}] +[\Sigma_S^{out}]}{2},
\label{eqSigmaS}
\end{equation}
where $[\Sigma_S^{in}]$ and $[\Sigma_S^{out}]$ are the in-scattering and the out-scattering matrices which model the rate of scattering of the electrons due to electron-phonon interactions \cite{QTDatta}. These scattering matrices are related to the electron ($[G^n$]) and the hole correlation ($[G^p]$) functions, given by
\begin{equation}
[\Sigma_S^{in}] = D_0\left( [G^n(E+h\omega_0) ]+ [G^n(E-h\omega_0)] \right ),
\label{eqSigmaSin}
\end{equation}
with
\begin{equation}
[\Sigma_S^{out}] = D_0 \left([G^p(E+h\omega_0)] + [G^p(E-h\omega_0)] \right ),
\label{eqSigmaSout}
\end{equation}
where $D_0$ is the scattering strength, and $\omega_0$ is the phonon frequency, which in the elastic case dealt with in this paper is taken to be zero. We focus on momentum and phase relaxation mechanisms \cite{QTDatta,LNEDatta} that are phenomenologically modeled via the the scattering strength $D_0$. Commonly, elastic interactions include phase breaking processes like momentum relaxation caused by acoustic phonons, electron-electron scattering and other possible near elastic processes. The above formalism of varying $D_0$ is also commonly used as a B\"uttiker probe style calculation to demonstrate the smooth transition from coherent transport to diffusive transport \cite{QTDatta,LNEDatta}, and serves the critical purpose of ``systematically'' adding such phase breaking processes over the coherent transport layer.
The electron and the hole correlation functions are again related to the electrons in-scattering and out-scattering matrices through the equations:
\begin{equation}
[G^n] = [G][\Sigma^{in}][G^{\dagger}], \quad [G^p] = [G][\Sigma^{out}][G]^{\dagger}.
\label{eqGnGp}
\end{equation}
The complete in-scattering and out-scattering functions are calculated as:
\begin{equation}
[\Sigma^{in}] = [\Gamma_H] f_H +[ \Gamma_C] f_C + [\Sigma_S^{in}],
\label{eqSigmain}
\end{equation}
and
\begin{equation}
[\Sigma^{out}] = [\Gamma_H] f_H + [\Gamma_C] f_C + [\Sigma_S^{in}],
\label{eqSigmaout}
\end{equation}
where $f_{H(C)}$ is the Fermi-Dirac distribution function of the hot (cold) contact and $\Gamma_{H(C)}$ represents the broadening matrix of the hot (cold) contact. The Equations \eqref{eqGS} -\eqref{eqSigmaSout} are calculated self-consistently with some initial guess of $\Sigma_S$.
The bias potential and the charging effect are encapsulated in the matrix $[U]$. The electronic charging effect is calculated using a self-consistent Poisson's equation in the transport direction $(\hat{z})$, given by
\begin{equation}
\frac{d^2}{dz^2}(U(z)) = \frac{-q^2}{\epsilon_o \epsilon_r} n(z),
\label{eqPoisson}
\end{equation}
where $q$ is the electronic charge, $\epsilon_{o(r)}$ is the free space (relative) permittivity. The electron density $n(z)$ along the transport direction is calculated as,
\begin{equation}
n(z) = \frac{1}{a} \int\frac{G^n(E)}{2 \pi} dE,
\label{eqnz}
\end{equation}
where $a$ is the unit length and $G^n$ is the diagonal element of energy resolved electron correlation function $[G^n(E)]$. The self-consistent solution of \eqref{eqGS}-\eqref{eqnz} is used to calculate the currents across two discrete device points $i$ and $i+1$ as
\begin{multline}
I_{i \rightarrow i+1} = \frac{iq}{h} \int dE [H(i,i+1)G^n(i+1,i) \\- G^n(i,i+1)H(i+1,i) ] ,
\label{eqJ}
\end{multline}
where, $A(i,j)$ represents the $(i,j)$ element of the matrix $[A]$ and
\begin{multline}
I_H^{Q} = \frac{2}{h} \int dE (E-\mu_H) \\ \times [H(i,i+1)G^n(i+1,i) \\- G^n(i,i+1)H(i+1,i) ] ,
\label{eqJH1}
\end{multline}
where $I_H^{Q}$ is the electronic heat current, originating from the hot contact . 
We can then evaluate the Seebeck coefficient as: 
\begin{equation}
S = \frac{-G_S}{G},
\label{eqS}
\end{equation}
where $G$ is the electrical conductance obtained from \eqref{eqI} by making $\Delta T=0$ at a small bias voltage that ensures the linear response operation. Similarly, from \eqref{eqI}, $G_S$ is obtained by setting $\Delta V=0$ at a finite $\Delta T$. Likewise from \eqref{eqIQ}, the quantity $G_P = I_Q/\Delta V$ is calculated by keeping $\Delta T=0$ and $G_Q=I_Q/\Delta T$ when $\Delta V=0$ are obtained. Now that we have all the transport coefficients the electronic thermal conductance is given by
\begin{equation}
G_K = G_Q - \frac{G_P G_S}{G}.
\label{eqGK}
\end{equation}
Using the above calculations of the linear response parameters via actual current calculations in the presence of scattering, we now proceed to analyze the thermoelectric performance of various superlattice structures. 
\section{Results and Discussion}
In the following section, we present the simulation results of superlattice TE performance parameters in the linear regime of operation. We will compare notes on how the elastic carrier scattering influences the TE parameters obtained from the coherent and non-coherent quantum thermoelectric transport methodology. In Tab.~\ref{tab:t1}, we enlist the parameter set for the calculations performed in this paper.
\subsection{Effect of Fermi Energy on TE Parameters}
\begin{figure}

        \subfigure[]{\includegraphics[height=0.18\textwidth,width=0.225\textwidth]{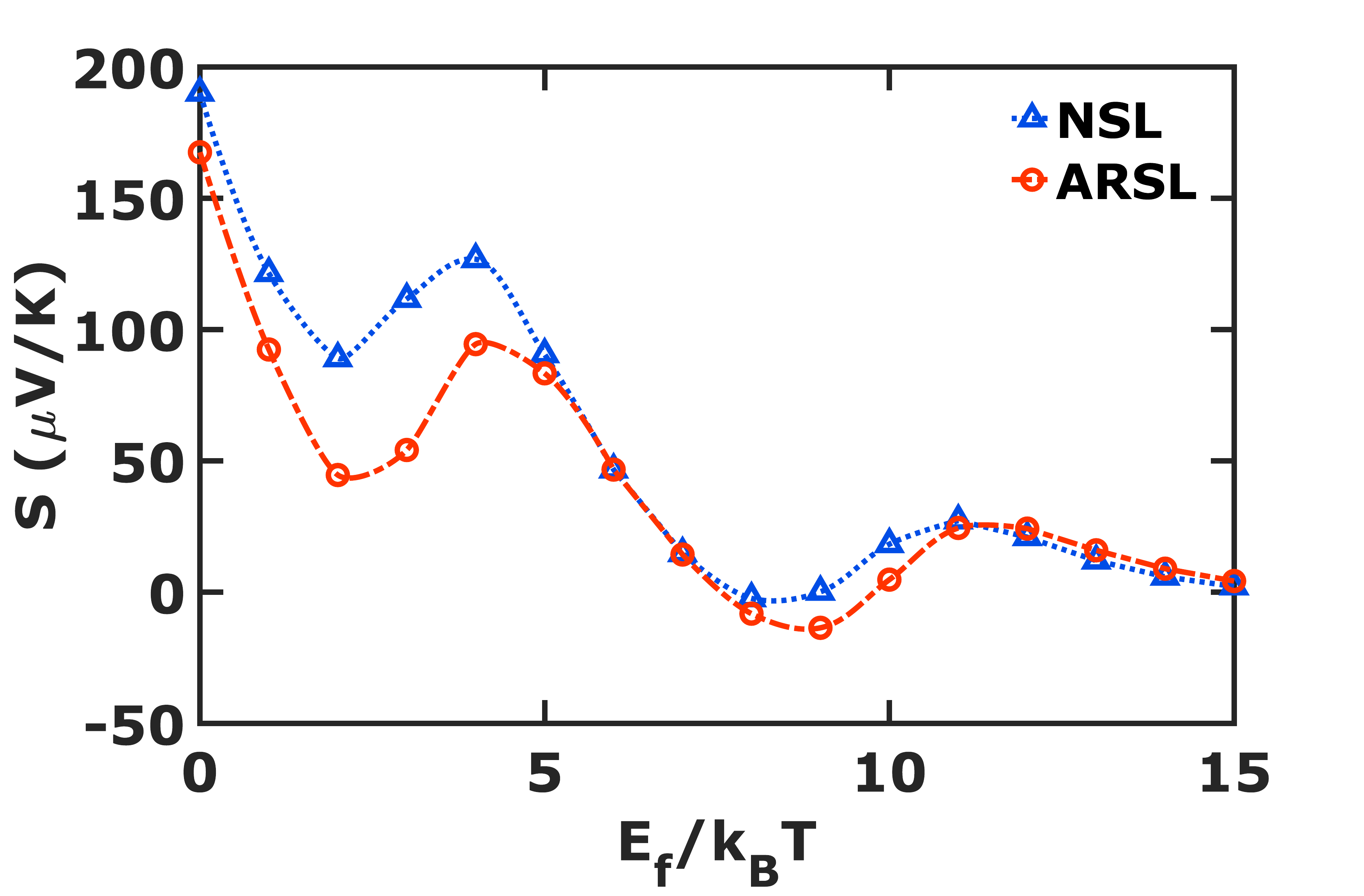}\label{NSL_ARSL_5B_S_Ef}}
        \quad
	\subfigure[]{\includegraphics[height=0.18\textwidth,width=0.225\textwidth]{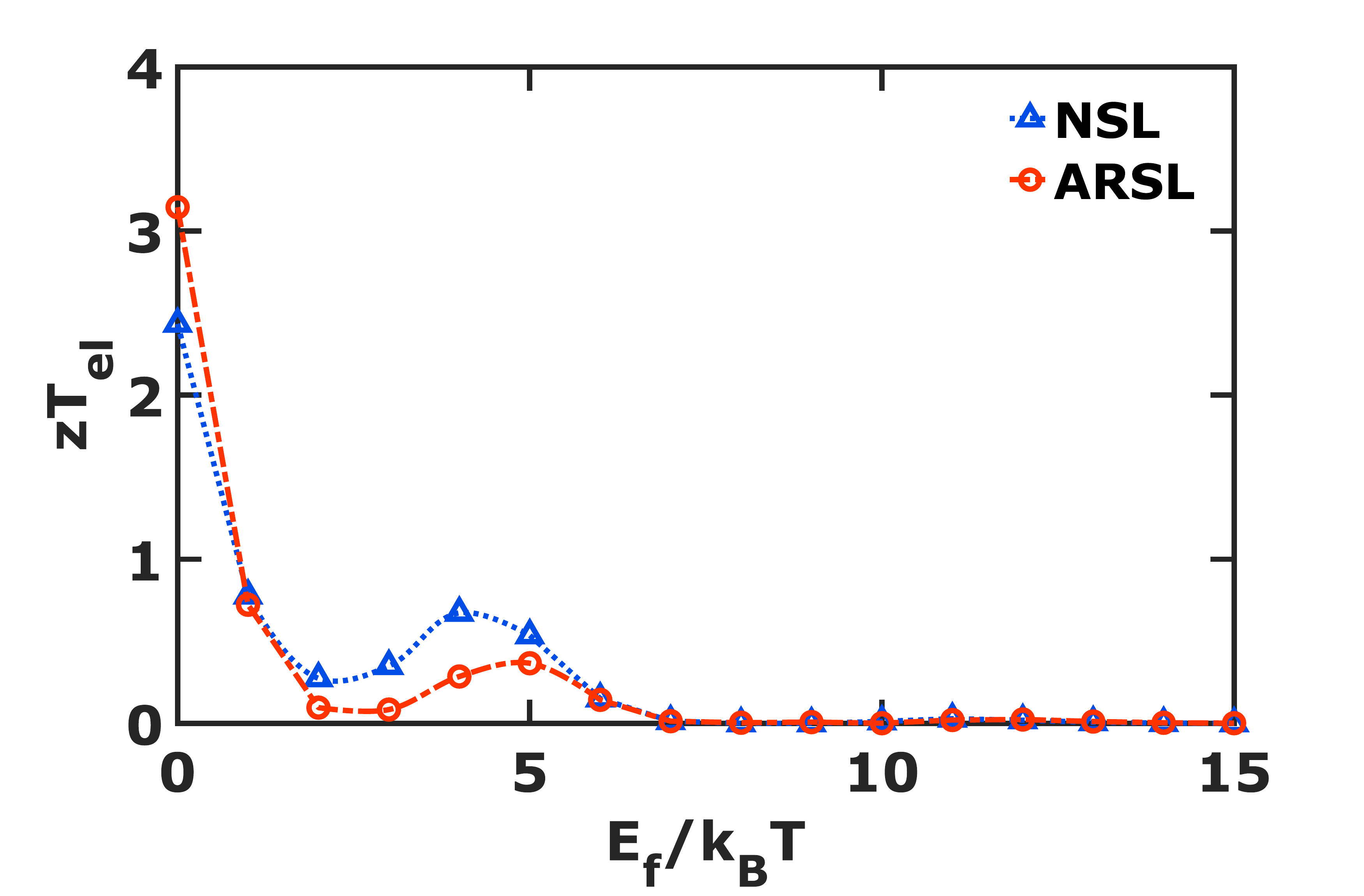}\label{NSL_ARSL_5B_ZT_Ef}}
	\caption{Color online: (a) Variation of the Seebeck coefficient as a function of $E_f$ for the 5-barrier NSL (blue curve) and ARSL (red curve) device configurations.
	(b) Figure of merit ($zT_{el}$) as a function of Fermi level $E_f$ for the 5-barrier NSL (blue curve) and ARSL (red curve) device configurations. Plots are sketched for the coherent transport simulation setup. The Fermi level $E_f$ is represented in the unit of $k_BT$.}
	\label{5B_ZT_S_Ef}
\end{figure}
\begin{figure}
\subfigure[]{\includegraphics[height=0.18\textwidth,width=0.225\textwidth]{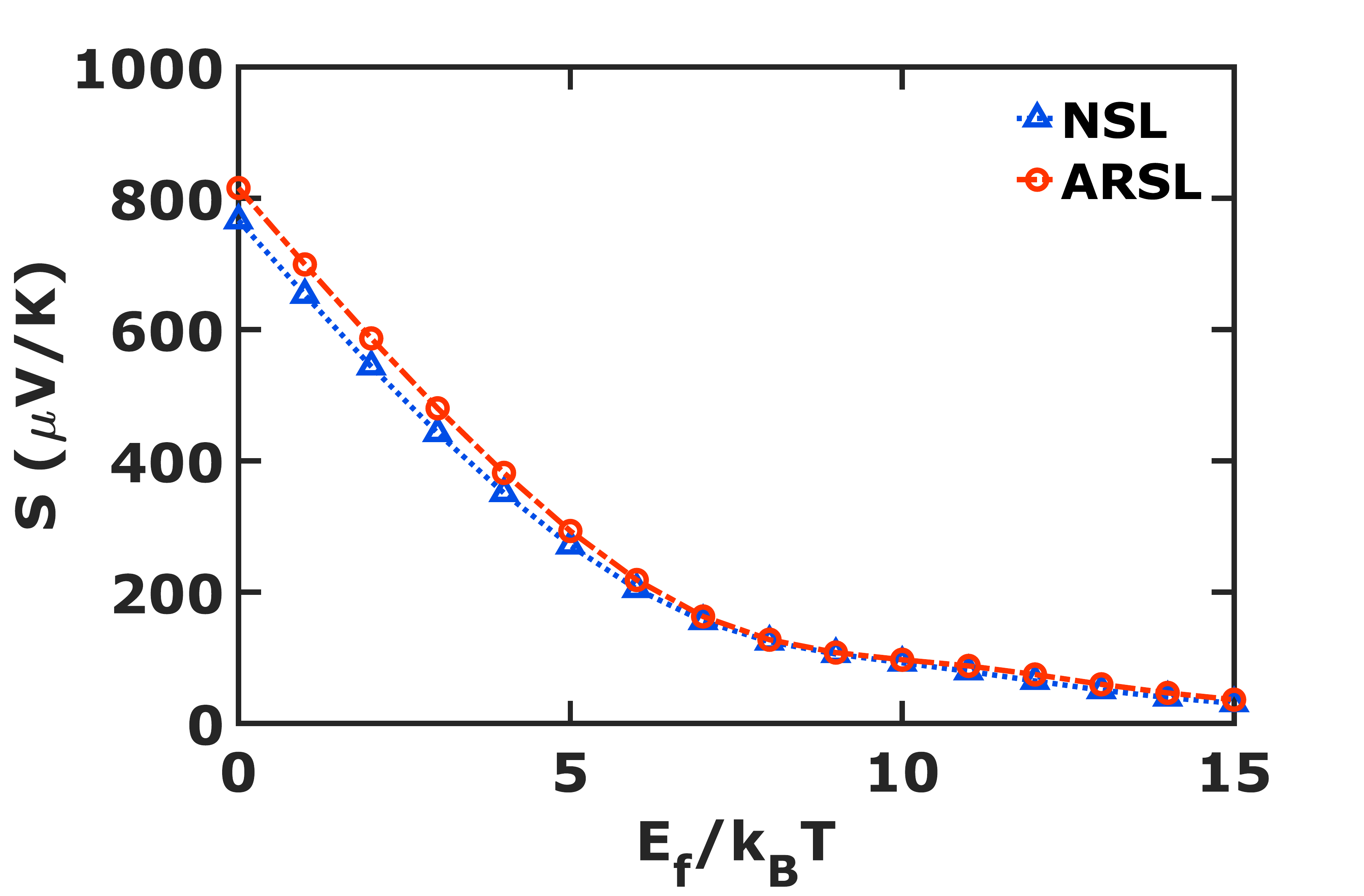}\label{NSL_ARSL_5B_S_Ef_Inelastic}}
	\quad
\subfigure[]{\includegraphics[height=0.18\textwidth,width=0.225\textwidth]{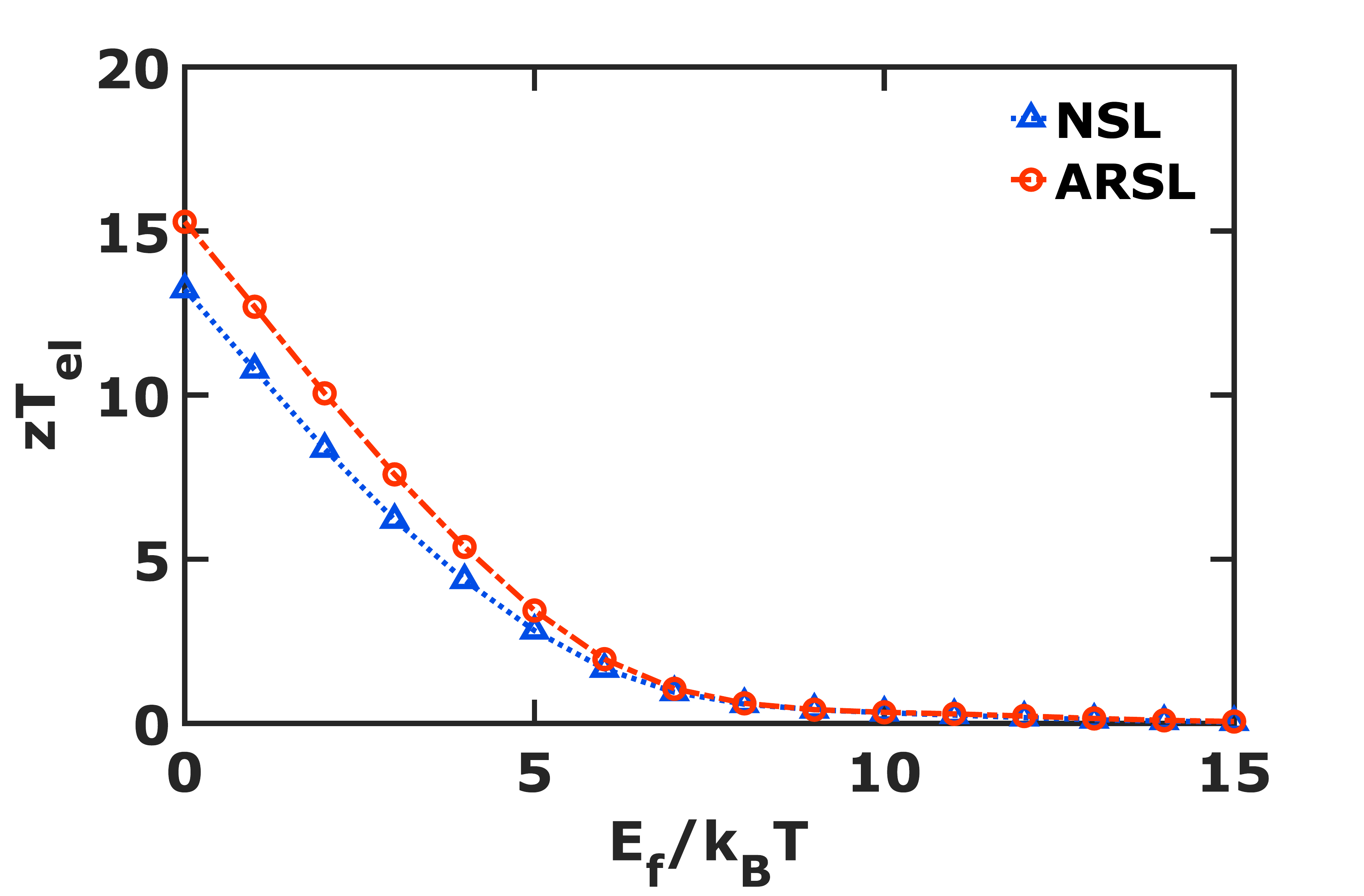}\label{NSL_ARSL_5B_ZT_Ef_Inelastic}}
		\caption{Color online: a) Variation of the Seebeck coefficient as a function of $E_f$ for the 5-barrier NSL (blue curve) and ARSL (red curve) device configurations. Here, the plots are sketched due to the non-coherent carrier transport solved via the elastic scattering simulation setup. (b) Figure of merit ($zT_{el}$) as a function of Fermi level $E_f$ for the 5-barrier NSL (blue curve) and ARSL (red curve) device configurations. ( The Fermi level $E_f$ is represented in the unit of $k_BT$.}
	\label{5B_ZT_Ef_Inelastic}
\end{figure}

Figure~\ref{5B_ZT_S_Ef} shows the variation of the Seebeck coefficient, $S$ and the electronic figure of merit, $zT_{el}$ as a function of channel Fermi level $E_f$, when the coherent transport simulation set up is active (as described in Fig.~\ref{Simulation_Flow}). For both the SL configurations (NSL \& ARSL), the maximum $zT_{el}$ is found at $E_f=0$, as shown in Fig.~\ref{NSL_ARSL_5B_ZT_Ef}, and thereafter it decreases. Likewise, the Seebeck coefficient is maximum at zero $E_f$, shown in Fig.~\ref{NSL_ARSL_5B_S_Ef}. This fact can be argued on the basis of the physics of thermoelectric transport, where the position of $E_f$ needs to below from the lower edge of transmission spectra \cite{Pankaj2019}, to achieve positive TE transport. In general, the position of $E_f$ governs the carrier concentration in the channel region, and thus on increasing $E_f$ (manually, in this work), we force the semiconductor to be degenerate. Although the electrical conduction is comparatively very close to the metal in a degenerate semiconductor, the thermopower ($S$) decreases on the same note. The Fermi level $E_f$ position is, therefore, an essential parameter in determining the TE performance of the concerned systems.\\
\begin{table}[t]
\centering
\caption{Parameter set used in the calculations}
\begin{tabular}{ |c|c|c|c| } 
		\hline
		SL & Parameters & Value & Unit \\ 
		\hline
		\hline
		1.) & $T_H$ & $330$ & Kelvin \\ 
		\hline
		2.) & $T_C$ & $300$ & Kelvin \\ 
		\hline
		3.) & $m_e$ & 0.07$m_o$ & $kg$ \\ 
		\hline
		4.) & $\delta E_c$ & 0.1 & $eV$ \\ 
		\hline
		5.) & $a$ (step size) & $3.0$ & $\AA$ \\ 
		\hline
		6.) & $D_o$ (scattering strength) & $0.01$ & $eV^2$ \\
		\hline 
		7.) & Well width & $6.0$ & $nm$ \\
		\hline 
		8.) & Barrier thickness & $4.0$ & $nm$ \\
		\hline 
	\end{tabular}
\label{tab:t1}
\end{table}
\indent We infer from the plots in Fig.~\ref{5B_ZT_S_Ef}, that there is not any practical advantage of using ARSL over NSL due to the inclusion of charging effect, as discussed in our previous work \cite{Pankaj2018}. The $zT_{el}$ reaches a maximum value of $3$ and $\approx2.4$ in the NSL and ARSL device configurations at zero $E_f$, respectively. Furthermore, at higher $E_f$, the NSL overtakes the ARSL configuration due to the mismatch of minibands and Fermi window at higher energies \cite{Pacher2001}. Thereby, we can say that by using the optimized superlattice structures, one can achieve a higher $zT_{el}$ as compared to the bulk TE counterpart. \\
\indent Now, we incorporate the elastic scattering in our calculations due to the phonon interaction and observe the variation in results obtained from coherent transport. In Figs.~\ref{NSL_ARSL_5B_ZT_Ef_Inelastic} \& \ref{NSL_ARSL_5B_S_Ef_Inelastic}, the Seebeck coefficient $S$ and $zT_{el}$ is plotted respectively as a function of $E_f$, by enabling the interaction block as described in Fig.~\ref{Simulation_Flow}, for the 5-barrier NSL and ARSL configurations. Unlike the plots in Fig.\ref{5B_ZT_S_Ef}, both quantities decrease gradually with the Fermi level, from the maximum value at $E_f=0$. Surprisingly, we achieve a maximum $zT_{el}$ of 13 and 15 at $E_f=0$ (shown in Fig.~\ref{NSL_ARSL_5B_ZT_Ef_Inelastic}) for the NSL and ARSL, respectively. This indicates that when we include the phonon interaction in electron transport, the TE parameters get improved. Likewise, the Seebeck coefficient enhances to a value of $760~\mu V/K$ in NSL and around $820~\mu V/K$ in the case of ARSL.
\subsection{Scaling Effect}
\begin{figure}
	\subfigure[]{\includegraphics[height=0.18\textwidth,width=0.225\textwidth]{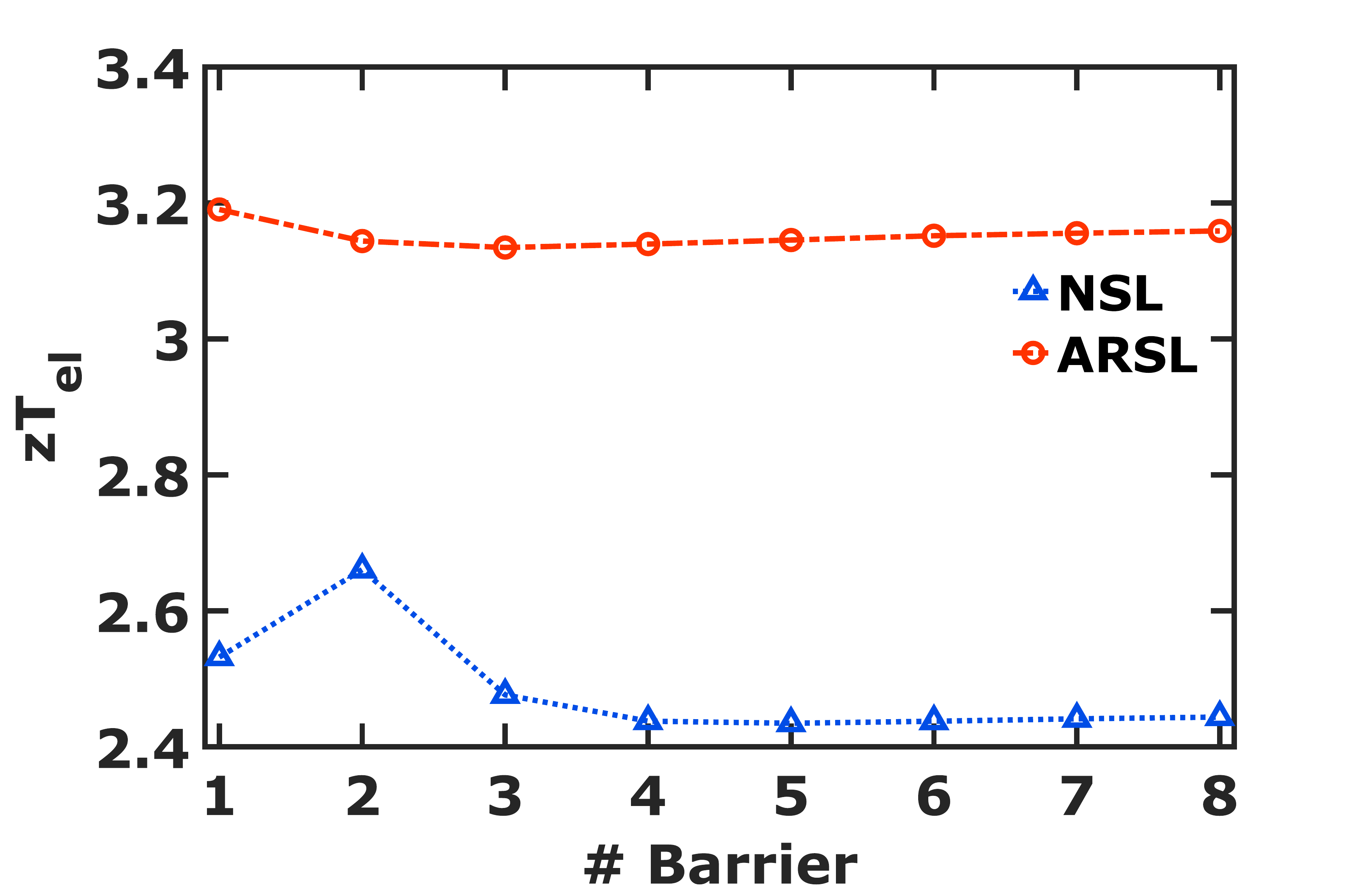}\label{NSL_ARSL_NB_ZT}}
	\quad
	\subfigure[]{\includegraphics[height=0.18\textwidth,width=0.225\textwidth]{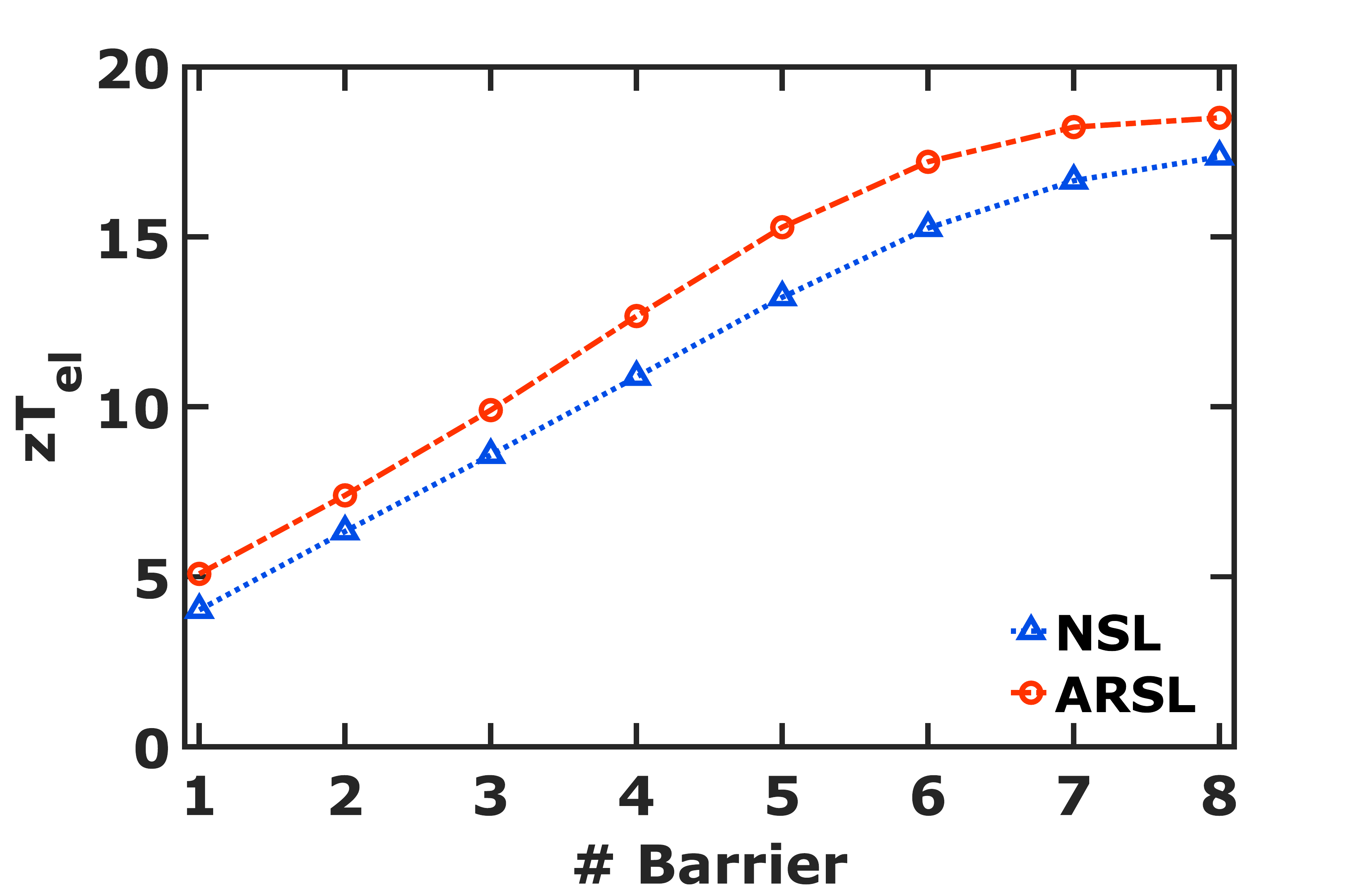}\label{NSL_ARSL_NB_ZT_Inelastic}}
	
	\caption{Color online: (a) Figure of merit ($zT_{el}$) as a function of number of barriers in the central channel region for the NSL (blue curve) and ARSL (red curve) device configurations when no interaction is subjected. (b) Similarly, $zT_{el}$ as a function of number of barriers for the NSL (blue curve) and ARSL (red curve) device configurations when elastic interaction is taken into account.}
	\label{NB_ZT}
\end{figure}

\begin{figure}
	\subfigure[]{\includegraphics[height=0.18\textwidth,width=0.225\textwidth]{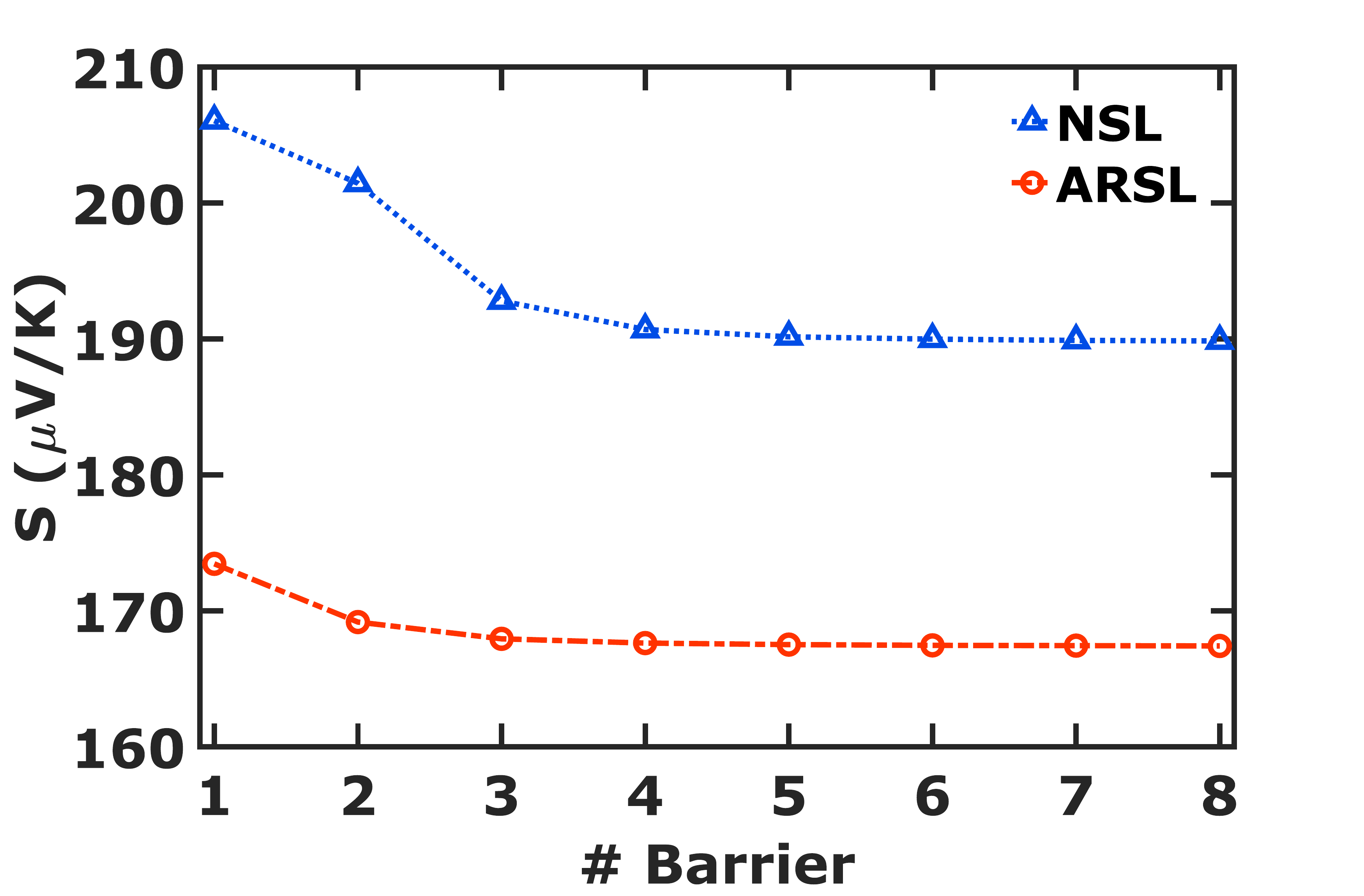}\label{NSL_ARSL_NB_S}}
	\quad
	\subfigure[]{\includegraphics[height=0.18\textwidth,width=0.225\textwidth]{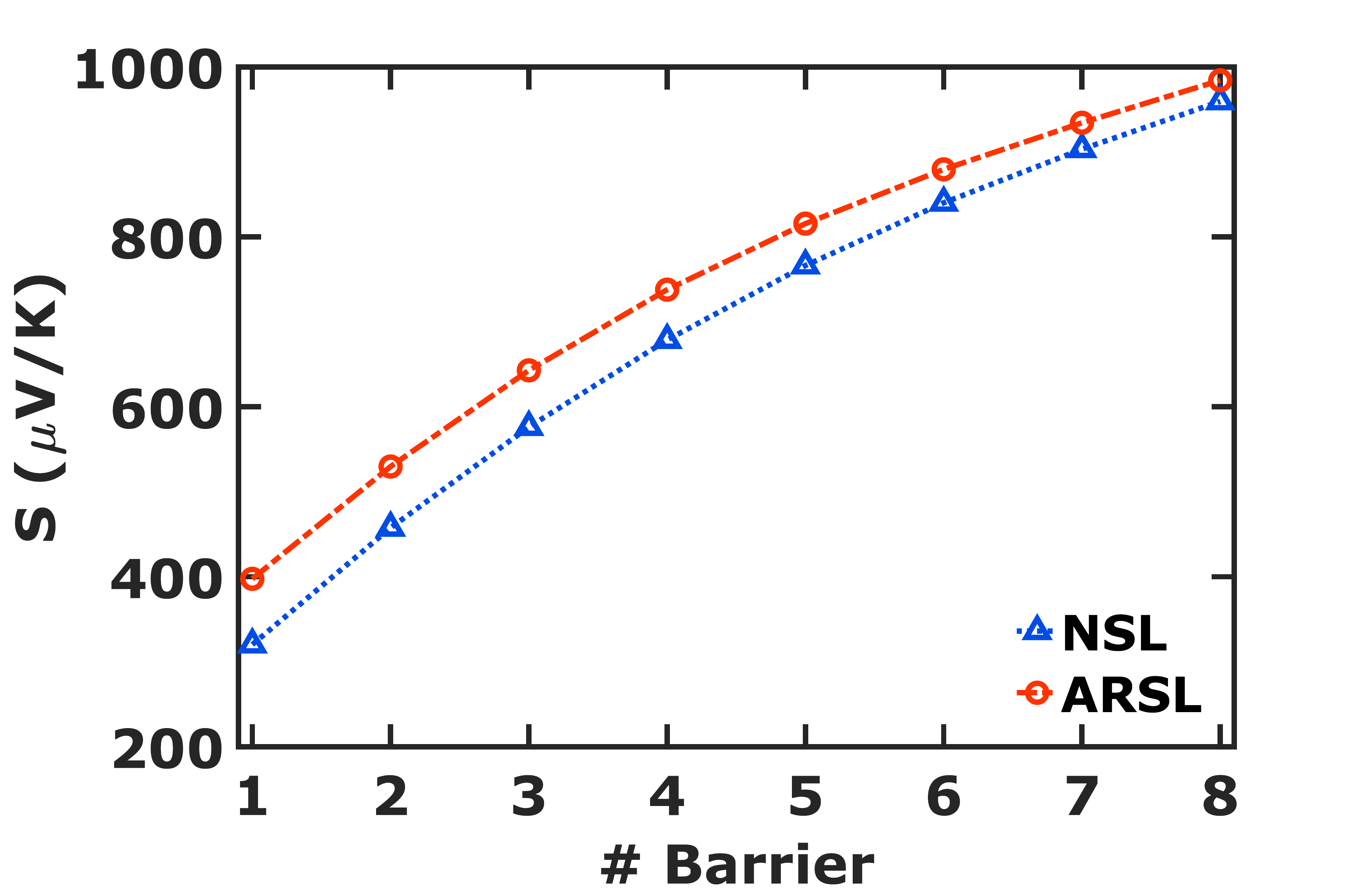}\label{NSL_ARSL_NB_S_Inelastic}}
	
	\caption{Color online: (a) Variation of Seebeck coefficient ($S$) as a function of number of barriers in the central channel region for the NSL (blue curve) and ARSL (red curve) device configurations when no interaction is subjected. (b) Similarly, $S$ as a function of number of barriers for the NSL (blue curve) and ARSL (red curve) device configurations when elastic interaction is taken into account.}
	\label{NB_S}
\end{figure}

\begin{figure}
	\subfigure[]{\includegraphics[height=0.18\textwidth,width=0.225\textwidth]{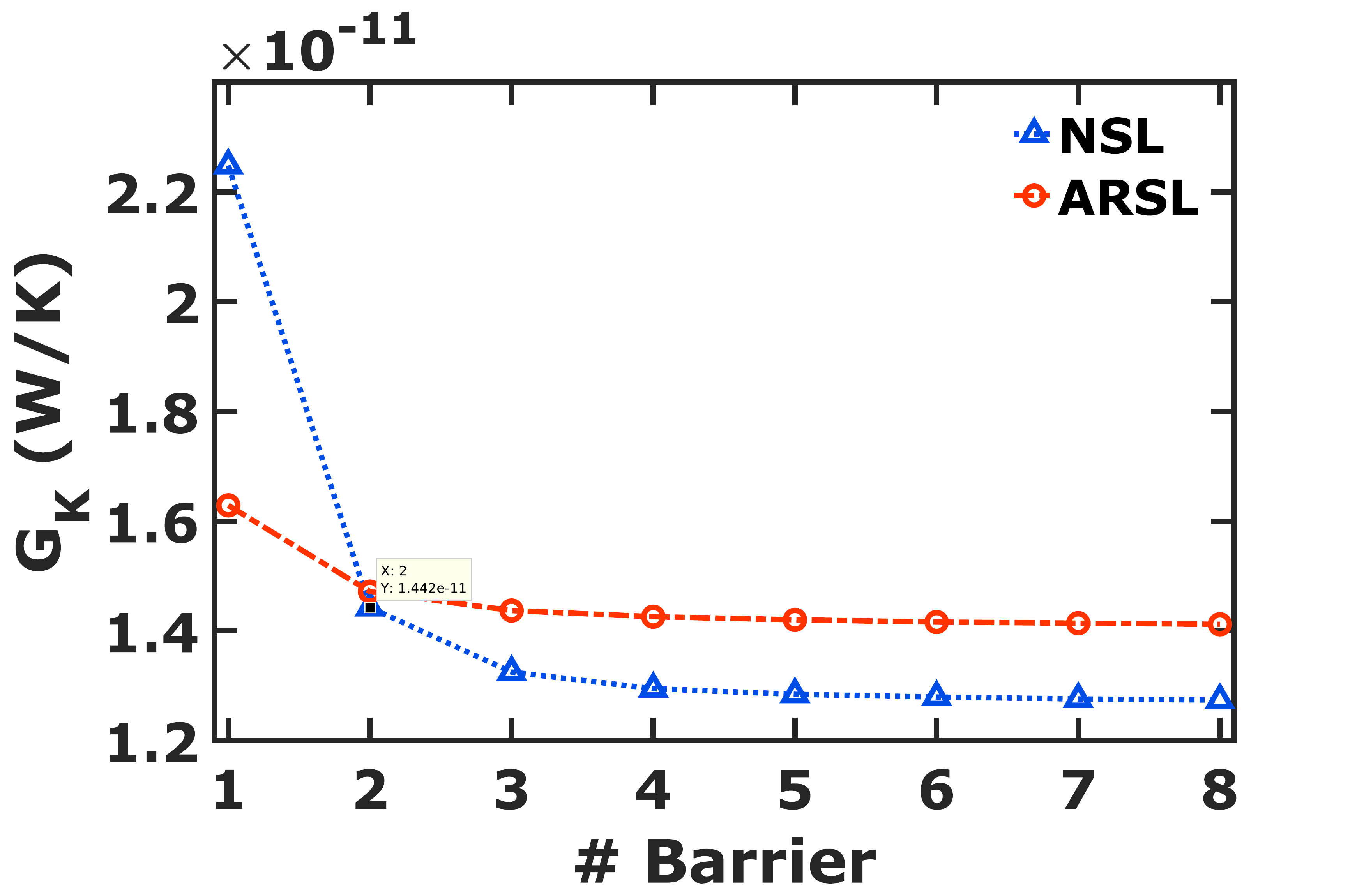}\label{NSL_ARSL_NB_GK}}
	\quad
	\subfigure[]{\includegraphics[height=0.18\textwidth,width=0.225\textwidth]{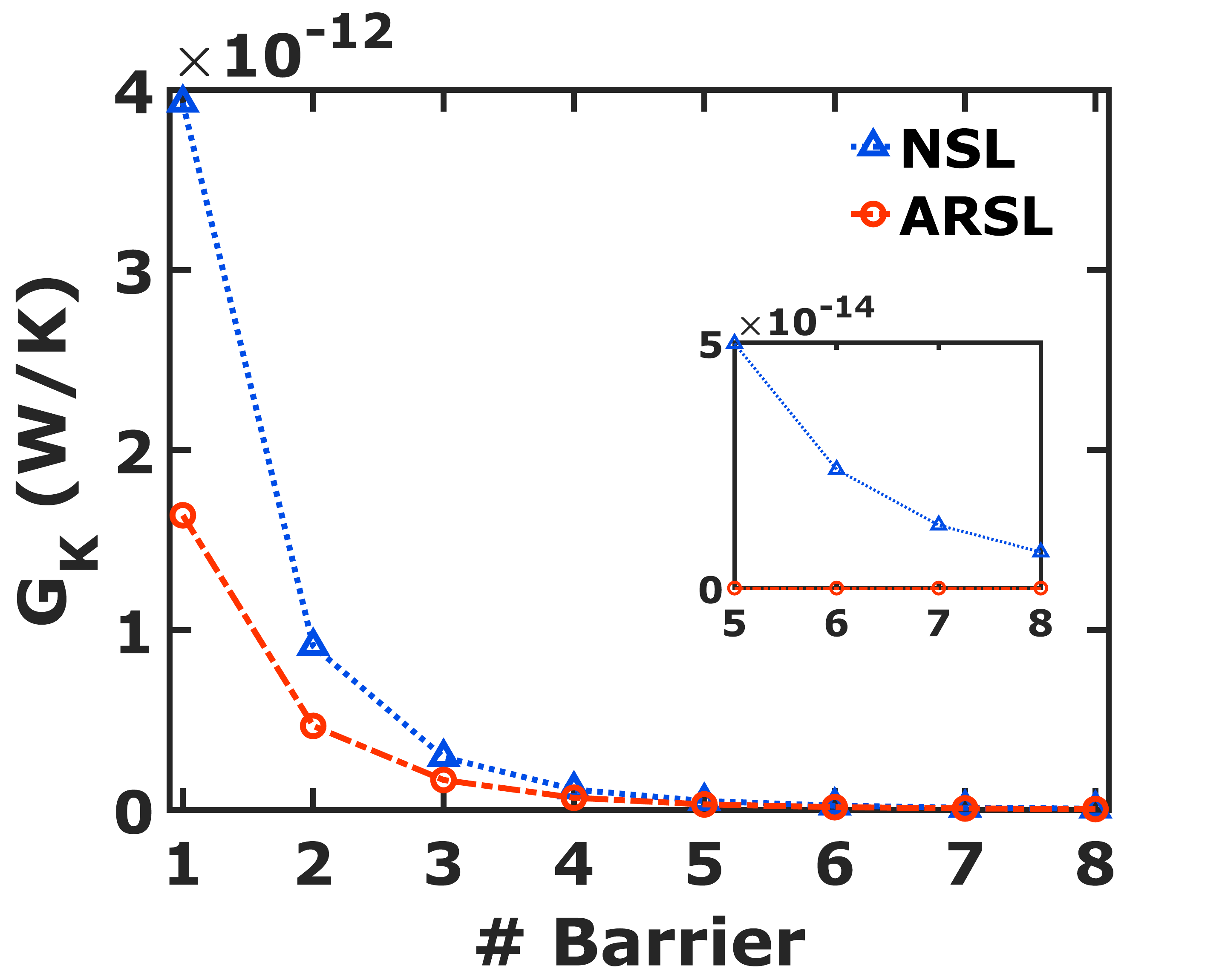}\label{NSL_ARSL_NB_GK_Inelastic}}
	
	\caption{Color online: (a) Thermal conductance ($G_K$) as a function of number of barriers in the central channel region for the NSL (blue curve) and ARSL (red curve) device configurations when no interaction is subjected. (b) Similarly, $G_K$ as a function of number of barriers for the NSL (blue curve) and ARSL (red curve) device configurations when elastic interaction is taken into account.}
	\label{NB_GK}
\end{figure}

To explore the effect of device length on the TE parameters, we vary the number of barriers in the central channel region of the device configurations keeping all the simulation parameters the same. In Fig.~\ref{NB_ZT}, we plot the figure of merit $zT_{el}$ as a function of the number of barriers for the coherent and non-coherent electron transport. 
We note from Fig.~\ref{NSL_ARSL_NB_ZT}, that the value of $zT_{el}$ increases to 2.7 for 2-barrier NSL configuration and thereafter saturates around 2.4, in the case of coherent transport. In the case of ARSL, $zT_{el}$ remains around 3.2, irrespective of the device length. We note here, that the use of anti-reflective layer in the superlattice enhances the figure of merit by 30\%, while ignoring the phonon interaction in electronic transport. The similar calculation approach, now enabling the elastic interaction block ($D_o\neq0$), is taken to plot the variation of $zT_{el}$ as a function of device length. From Fig.~\ref{NSL_ARSL_NB_ZT_Inelastic}, we notice that the figure of merit $zT$ gets improved in both the cases, contrarily when $D_o=0$. The parameter $zT_{el}$ gradually increases with the device length, but slightly higher in the ARSL configuration. The maximum numerical value of $zT_{el}$ we obtain here is approximately 17 \& 18 in 8-barrier NSL and ARSL configurations, respectively. Beyond this device length, the value gets saturated. This can be understood by the fact that when we increase the number of barriers in the channel region, the mean free path of phonons gets diminished \cite{Sofo-Mahan1994, Broido1995, Balandin2003} and thus doesn't affect the flow of electrons through the channel. \\
\indent Figure~\ref{NB_S} shows the variation of the Seebeck coefficient as a function of number of barriers for the one dimensional transport calculation. From Fig.~\ref{NSL_ARSL_NB_S}, it is evident that the value of $S$ in NSL configuration is better than the ARSL configuration when no interaction is included. But, taking the elastic scattering into account, the Seebeck coefficient $S$ enhances in both device structures, shown in Fig.~\ref{NSL_ARSL_NB_S_Inelastic}. Here, we get an advantage in the case of ARSL configuration and thus reaches up to a value of $S=1100~\mu V/K$ at 8-barriers. Generically, there is a reduction in the electronic conduction as we increase the barriers in the channel; however, the Seebeck coefficient increases as mathematically observed from the \eqref{eqS}. Thus, there is a trade-off between the electrical conductance $G$ and the Seebeck coefficient $S$, when the device is forced to operate in the non-linear regime.\\
\indent The above analysis, so far, can also be supported by the conduction of heat carried out by the electrons in the device. While lattice phonons also contribute to the thermal conduction, the superlattice interfaces and boundaries affect its flow \cite{GangChen2005}. Here, we calculate the electronic part of the thermal conduction, as explained in the \eqref{eqGK}, and demonstrate that such interfaces also affect the electronic heat flow in the presence of elastic scattering. In Fig.~\ref{NB_GK}, the variation of thermal conduction ($G_K$) as a function of number of barriers is shown. On comparing the plots from Figs.~\ref{NSL_ARSL_NB_GK} \& \ref{NSL_ARSL_NB_GK_Inelastic}, the thermal conduction ($G_K$) is greatly reduced by a factor of 10 when we include the interaction part in the calculations. We also note from Fig.~\ref{NSL_ARSL_NB_GK_Inelastic} that the $G_K$ value is almost zero, however in the case of ARSL, it approaches as low as (shown in the inset of Fig.~\ref{NSL_ARSL_NB_GK_Inelastic}), for the 8-barrier device length. The optimum layer thickness and stoichiometry proportion of the superlattice structures shrink the mean free path or wavelength of the heat carriers (i.e. phonons in semiconductor) imposes an additional resistance to the thermal transport and thus reduces the thermal conduction.\\
\begin{figure}
	\subfigure[]{\includegraphics[height=0.18\textwidth,width=0.225\textwidth]{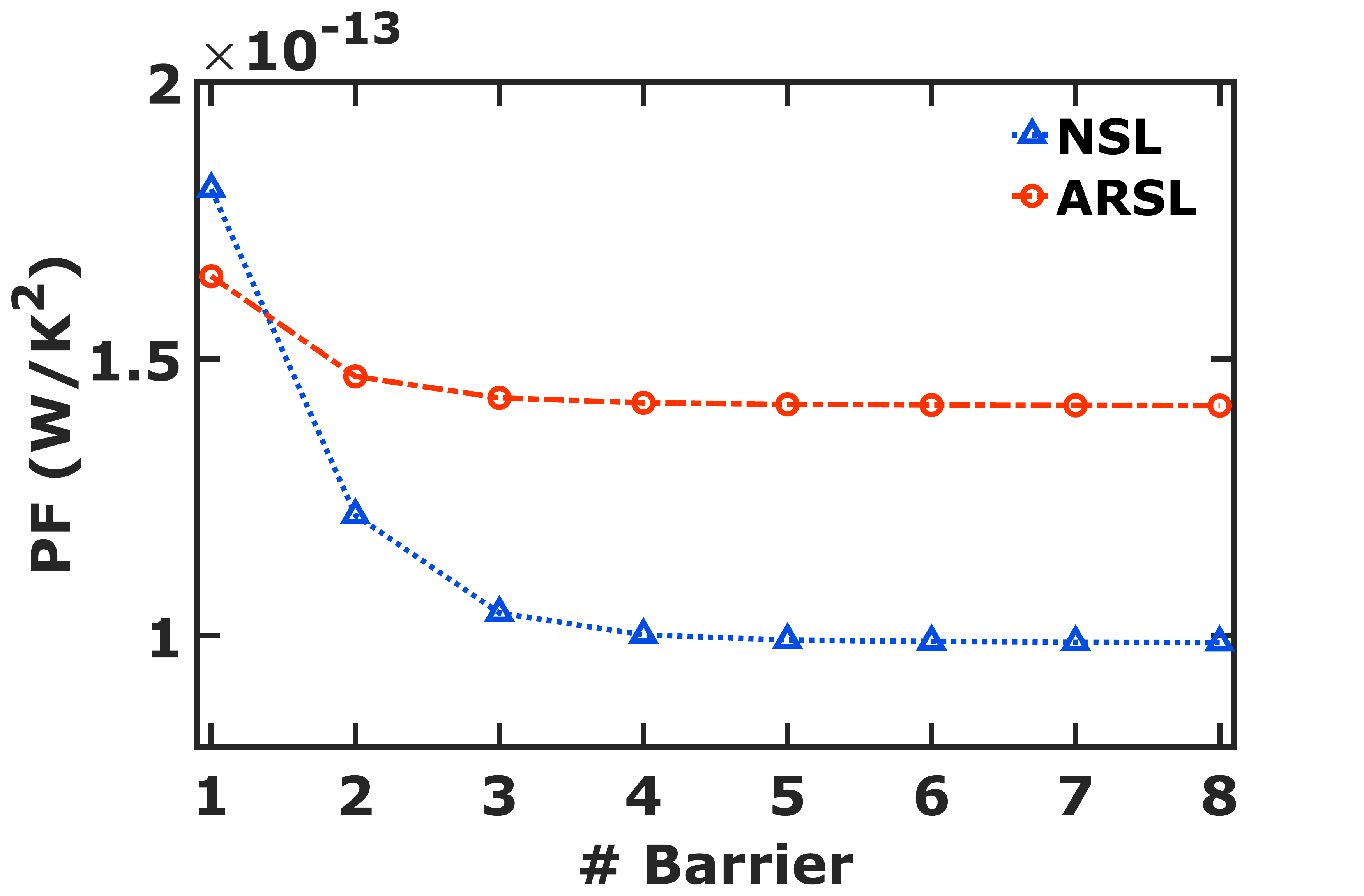}\label{NSL_ARSL_NB_PF}}
	\quad
	\subfigure[]{\includegraphics[height=0.18\textwidth,width=0.225\textwidth]{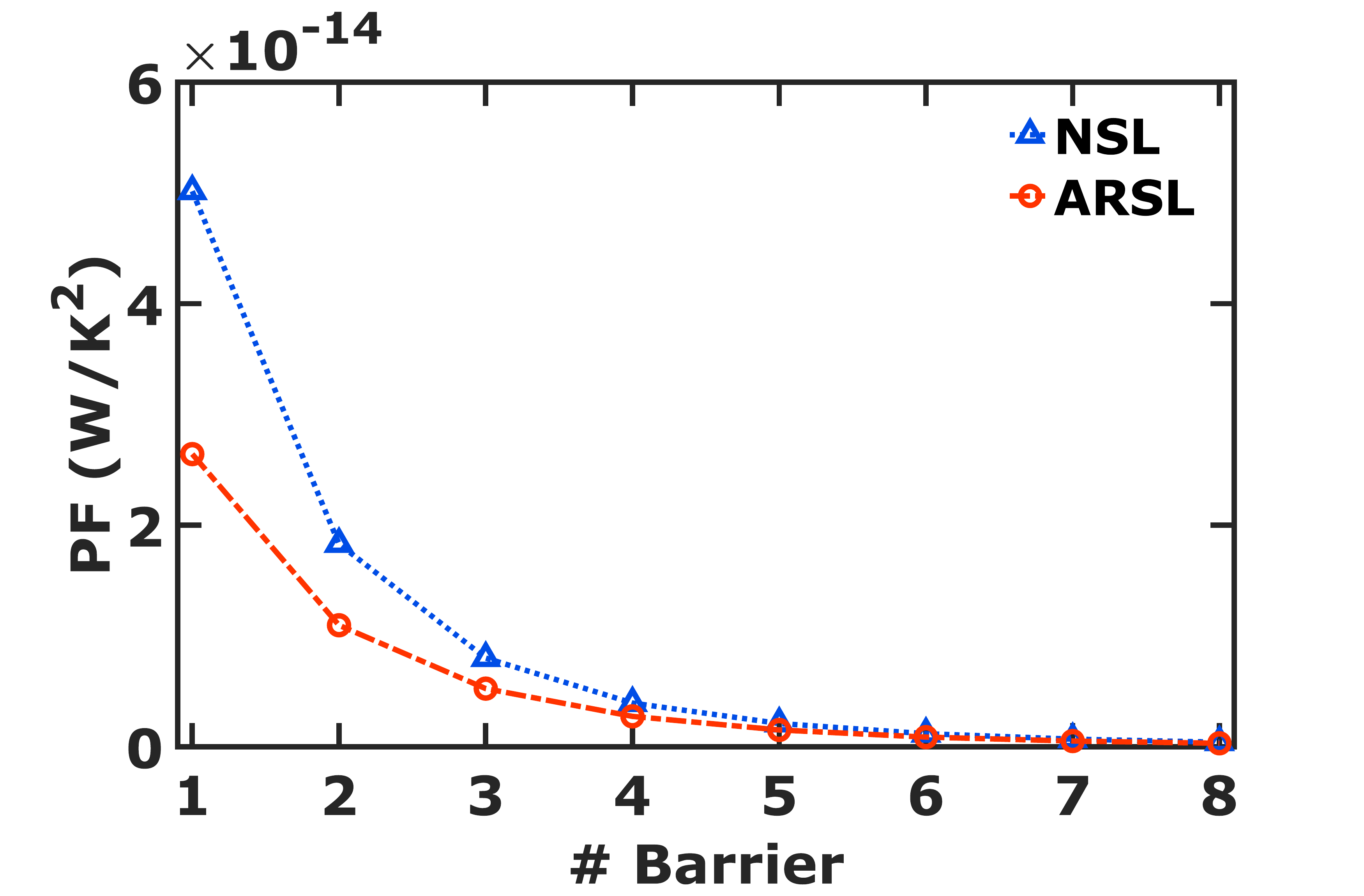}\label{NSL_ARSL_NB_PF_Inelastic}}
	
	\caption{Color online: (a) TE generator power factor ($PF$) as a function of number of barriers in the central channel region for the NSL (blue curve) and ARSL (red curve) device configurations when no interaction is subjected. (b) Similarly, the $PF$ as a function of number of barriers for the NSL (blue curve) and ARSL (red curve) device configurations when elastic interaction is taken into account.}
	\label{NB_PF}
\end{figure}
\indent To benchmark a TE material or device structure, the quantity power factor (PF$=S^2G$) that defines a practical application of the TE power generator. For a functional power generator, one has to look for the maximum power factor. However, it is challenging to enhance PF without disturbing the other thermoelectric coefficients, as explained via the inconsistency between $S$ and $G$. Although it is suggested that in a superlattice where carriers are confined within a thin layer (here in the transport direction $\hat{z}$) one can enhance the PF \cite{Hicks1993-1, Hicks1993-2}. We plot the variation of PF as a function of number of barriers in Figs.~\ref{NSL_ARSL_NB_PF} \& \ref{NSL_ARSL_NB_PF_Inelastic} for the coherent and non-coherent simulation setups, respectively. In both the SL configurations, the value of PF decreases with the number of barriers. Similarly, when the elastic scattering is taken into account the PF value is reduced by a factor of 10. This could be understood by rationally observing the reduction of electrical conduction in the lengthy device. \\
\indent The results, particularly those of Fig. ~\ref{NB_GK} and Fig. ~\ref{NB_PF}, elucidate the role of elastic scattering in the drastic reduction of electronic thermal conductance, albeit a decrease power factor as a result of a corresponding decrease in electrical conductance. Coupled with the increase in Seebeck coefficient due to electronic filtering effects of the "box car" transmission function, our results demonstrate the interplay of coherent quantum effects dictated by electronic filtering along with scattering effects as a route toward electronic engineering of thermoelectric structures. While the presence of interfaces is known to kill phonon thermal conduction, our analysis shows that non-coherent processes in superlattice structures can effectively kill electronic thermal conduction also. 
\section{Conclusion}
Using the {\it{dissipative}} non-equilibrium Green's function formalism coupled self-consistently with the Poisson's equation, we reported an enhanced figure of merit $zT$ in the multi-barrier device designs. The proposed enhancement is a result of a drastic reduction in the electronic thermal conductance triggered via non-coherent transport. We show that a maximum $zT$ value of 18 can been achieved via the inclusion of non-coherent elastic scattering processes in the electron transport. There is also a reasonable enhancement in the Seebeck coefficient, with a maximum of $1000~\mu V/K$, which we attribute to an enhancement in electronic filtering arising from the non-coherent transport. Distinctly the thermal conduction is drastically reduced as the length of the superlattice scales up, although the power factor shows an overall degradation. We believe that the analysis presented here could set the stage to understand better the interplay between non-coherent scattering and coherent quantum processes in the electronic engineering of heterostructure thermoelectric devices.\\
{\it{Acknowledgements:}} The Research and Development work undertaken in the project under the Visvesvaraya Ph.D. Scheme of Ministry of Electronics and Information Technology, Government of India, is implemented by Digital India Corporation (formerly Media
Lab Asia). This work was also supported by the Science and
Engineering Research Board (SERB), Government of
India, Grant No. EMR/2017/002853 and Grant No. STR/2019/000030.

\bibliographystyle{apsrev}
\bibliography{Reference}

\end{document}